\documentclass[superscriptaddress,pre,english,reprint]{revtex4-1}
\usepackage{amsmath}
\usepackage{bm}
\usepackage{graphicx}
\usepackage{hyperref}

\begin{document}

\title{Influence of atomic kinetics on inverse bremsstrahlung heating and nonlocal thermal transport} 

\author{Hai P. Le}
\thanks{Corresponding author}
\email{hle@llnl.gov}
\author{Mark Sherlock}
\author{Howard A. Scott}
\affiliation{Lawrence Livermore National Laboratory, Livermore, California 94551, USA}


\date{May 1, 2019}
 
\begin{abstract}
This paper describes a computational model that self-consistently combines physics of kinetic electrons and atomic processes in a single framework. The formulation consists of a kinetic Vlasov-Boltzmann-Fokker-Planck equation for free electrons and a non-Maxwellian collisional-radiative model for atomic state populations. We utilize this model to examine the influence of atomic kinetics on inverse bremsstrahlung (IB) heating and nonlocal thermal transport. We show that atomic kinetics affects non-linear IB absorption rates by further modifying the electron distribution in addition to laser heating. We also show that accurate modeling of nonlocal heat flow requires a self-consistent treatment of atomic kinetics, because the effective thermal conductivity strongly depends on the ionization balance of the plasma.
\end{abstract}


\maketitle 


\section{Introduction}
\label{sec:intro}
The interplay between kinetic physics of electrons and atomic processes has previously been identified to play an important role in laser-produced plasmas~\cite{langdon_nonlinear_1980,ditmire_time-resolved_1998,hansen_hot-electron_2002,robinson_fast_2006,medvedev_short-time_2011,hau-riege_nonequilibrium_2013}. However, an adequate modeling treatment is often out of reach due to high computational cost. Kinetic modeling of electrons itself presents a challenge due to the high dimensionality of the governing equation. 
This complexity is exacerbated in mid- to high-$Z$ nonequilibrium plasmas, where determining the ion charge state that feeds into the electron kinetic equation may require solving atomic kinetics rate equations for hundreds or thousands of states and tens of thousands of transitions.

Most electron kinetic models assume a fixed ionization state and neglect the effects of atomic kinetics. This assumption is only valid for a fully ionized or local thermodynamic equilibrium (LTE) plasma. These condititions, however, are rare in mid- to high-Z laboratory laser-produced plasmas. These plasmas are often not fully ionized. Even if the temperature is high enough for that to occur, it still takes a finite amount of time for the system to ionize such that atomic kinetics must be accounted for in the model. In addition, energy losses occurring via transport and radiative processes (due to finite size and spatial gradients) also prevent the system from staying fully ionized. In either case, atomic kinetics plays an important role in determining the ionization balance, which affects other processes through absorption, transport, radiative properties, etc. Atomic kinetics is also sensitive to the electron distribution, because the cross sections for these processes can vary significantly with respect to the impact energy. In the present study, we will examine the influence of atomic kinetics on two basic problems relevant to laser-produced plasmas: inverse Bremsstrahlung (IB) heating and nonlocal thermal transport.

The rest of the paper is organized as follows. In Sec. \ref{sec:model}, we introduce a framework to self-consistently model atomic kinetic processes and nonthermal (kinetic) electrons. The two main components are a kinetic equation for the electrons and a non-Maxwellian collisional-radiative model for atomic kinetics. The influence of atomic kinetics on IB heating and nonlocal thermal transport are examined in Sec. \ref{sec:ib} and Sec. \ref{sec:transport}. Finally, a summary is given in Sec. \ref{sec:summary}. Some details of the atomic kinetics treatment and IB absorption are described in the Appendix.

\section{Computational Model}
\label{sec:model}
A kinetic model is used to describe the time evolution of the electron distribution function coupled with ions via elastic and inelastic processes. The resultant Vlasov-Boltzmann-Fokker-Planck equation (VBFP) for the electron reads:
\begin{equation}
\label{eq:vbfp}
\partial_t f + \bm{v} \cdot \nabla f + \frac{e}{m_e} \bm{E} \cdot \nabla_v f = \mathcal{C}(f) + \mathcal{Q}(f)
\end{equation}
where $f \equiv f(\bm{x},\bm{v},t)$ is the electron velocity distribution function, and $\mathcal{C}$ and $ \mathcal{Q}$ are the elastic and inelastic collision operators, respectively. The name Vlasov-Boltzmann-Fokker-Planck comes from the fact that Eq. (\ref{eq:vbfp}) contains two Vlasov terms (last two terms on left hand side), a Fokker-Planck term ($\mathcal{C}$) and a Boltzmann term ($\mathcal{Q}$). When atomic kinetics is not considered, $\mathcal{Q}$ is absent and we end up with a Vlasov-Fokker-Planck (VFP) model. The elastic collision operator $\mathcal{C}$ includes both electron-electron (ee) and (elastic) electron-ion (ei) collisions, i.e., $\mathcal{C}(f) = \mathcal{C}_{ee}(f) +\mathcal{C}_{ei}(f)$. Both $\mathcal{C}_{ee}$ and $\mathcal{C}_{ei}$ depend on atomic kinetics because the collision frequencies scale with the ionization of the plasma, i.e., $\nu_{ee} \sim N_e$ and $\nu_{ei} \sim Z N_e$ where $N_e = \int_{-\infty}^\infty f(\bm{v}) \, d\bm{v}$.

The inelastic collision operator $\mathcal{Q}$ is responsible for the coupling between free electrons and the atomic state distribution. In this work, inelastic processes refer to all atomic kinetic processes that affect the atomic state distribution and ionization balance. They consist of a large number of processes that directly depend on atomic state densities, e.g., collisional excitation/deexcitation, ionization/recombination, autoionization/electron capture, photo-ionization and radiative recombination. Detailed expressions for these terms are given in appendix \ref{app:inelas}. Time-dependent Boltzmann kinetic model had been used in the past to simulate ultrashort-pulse high intensity laser experiments (see, for example, \cite{abdallah_time-dependent_2003,varga_non-maxwellian_2013,gao_ultrafast_2017}). These studies often assume a spatially homogenous (zero-dimensional) plasma to simplify the calculation.

Due to complexity and high computational cost, most electron kinetic models (with two exceptions to be mentioned later) neglect $\mathcal{Q}$ and further assume that the plasma ionization state $Z$ is fixed in time~\cite{epperlein_two-dimensional_1988,weng_inverse_2009,thomas_review_2012,tzoufras_multi-dimensional_2013,sherlock_comparison_2017}. Strictly speaking, this assumption is only valid for a fully ionized plasma where there are not any inelastic processes occurring. When the atom is partially filled with bound electrons, inelastic processes can happen, and $\mathcal{Q}$ cannot be safely neglected in general. The main reason is because $\mathcal{C}_{ee} = 0$ is a necessary condition for $\mathcal{Q} = 0$, in which case both electrons and atomic states must follow their equilibrium distributions; this is known as the local thermodynamic equilibrium (LTE) limit. Therefore, the plasma is always in non-LTE ($\mathcal{Q} \neq 0$) when $\mathcal{C}_{ee} \neq 0$, and the atomic state distribution cannot simply be determined from equilibrium (Saha-Boltzmann) distributions. It is possible, however, to have $\mathcal{C}_{ee} = 0$ when $\mathcal{Q} \neq 0$, in which case the free electrons are thermalized. Most non-LTE atomic kinetics models operate based on this assumption since all transition rates involving free electrons can be conveniently parameterized in terms of $n_e$ and $T_e$, instead of the full distribution function $f$~\cite{ralchenko_modern_2016}.

It is not always clear how $\mathcal{Q}$ affects the distribution (and other physical processes that depend on the shape of the distribution, e.g., transport). In general, inelastic processes can play two competing roles: ionization and excitation from ground state (or transitions involving electrons from inner-shell) with large transition energies (compared to the average energy of the electron) can cause depletion of the high energy tail (dethermalization), while excitation/deexcitation collisions between high-lying states, due to their small transition energies, act more like elastic collisions that thermalize the distribution. We note that since excitation/deexcitation does not change the total electron density, $\mathcal{Q}$ can be nonzero even when $Z$ is constant in time.

We mention two previous studies which included ionization kinetics in VFP codes. The first one is due to Town et al.~\cite{town_fokker-planck_1995} In this work, the authors coupled their VFP code to a simple ionization kinetics model to simulate short pulse laser-solid interaction. The ionization model was highly simplified that it only took in account collisional ionization and recombination; other processes (such as excitation, auger ionization, radiative recombination) are neglected. Either and Matte~\cite{ethier_electron_2001} also attempted to include atomic kinetics by coupling an average atom model to their VFP code. Their average atom model gave a slightly better description of the atomic kinetics, compared to Town et al., since it took into account excitations between bound states. However, the accuracy of an average atom model is severely limited, particularly for mid- to high-Z elements due the lack of an adaquate treatment for autoionization and dielectronic recombination (see, for example, a discussion in \cite{scott_advances_2010}).

In this work, we employ a collisional-radiative model to simulate non-LTE atomic kinetics~\cite{scott_cretin-radiative_2001,chung_flychk:_2005,hansen_superconfiguration_2011,ralchenko_modern_2016}. The distribution of atomic populations is calculated by solving a set of coupled rate equations:
\begin{equation}
\label{eq:cr}
\frac{ d \bm{y} } {dt} = \bm{A} \cdot \bm{y}
\end{equation}
where $\bm{y}$ is the vector of atomic state densities and $\bm{A}$ is the rate matrix. All inelastic processes described in the previous paragraph are included in $\bm{A}$. Eq (\ref{eq:cr}) is coupled to (\ref{eq:vbfp}) via the rate matrix, because transition rates involving free electrons directly depend on the distribution function $f$. For example, an excitation (or ionization) rate from $i$ to $j$ is obtained from $\int f \, \sigma_{i\rightarrow j}v \, d \bm{v}$ where $\sigma_{i\rightarrow j}$ is the collision cross section. Radiative excitation/deexcitation rates are included in the rate matrix $\bm{A}$ but not in $\mathcal{Q}$ because they do not change the electron distribution.

The numerical solution of the VBFP equation is based on the KALOS formalism \cite{bell_fast_2006}. The VFP part of the model is from the VFP code K2 \cite{sherlock_comparison_2017}, which also includes a self-consistent treatment of the electric field. The distribution function is expanded in spherical harmonics in velocity space, i.e., $f (\bm{x}, \bm{v}, t) = \sum_{n=0}^{n_{\max}} \sum_{m=-n}^{n} f_n^m (\bm{x}, v, t) Y_n^m (\theta, \varphi) $, and the expansion coeffcients are evolved in time according to the kinetic equation. All simulations performed in this study are at most one-dimensional in space, and the expansion is truncated at $n_{\max} = 1$. Since there is no magnetic field, we only need to evolve two coefficients, i.e., $f_0^0$ and $f_1^0$. For clarity, we now drop the superscript and refer to them as $f_0$ and $f_1$. We also assume that inelastic processes do not contribute significantly to momentum scattering, i.e., $\mathcal{Q}$ has no effect on $f_1$ and is only applied to $f_0$. Advection, acceleration and collisions are handled separately using operator-spliting techniques. While advection is treated explicitly (for efficient parallelization), acceleration and collisions are treated implicitly. The collision terms in $\mathcal{Q}$ are discretized using a conservative formulation detailed in~\cite{le_conservative_2017}. Eqs. (\ref{eq:vbfp}) and (\ref{eq:cr}) are solved self-consistently at each time step to update both the electron distribution $f$ and the atomic state distribution $\bm{y}$. Radiation transport is neglected, as the plasma is assumed to be optically thin to all radiative transitions. 

The atomic physics data used in this work is constructed based on the screened-hydrogenic model. The energy levels are described by principal quantum numbers (pqn), including single and double excitations from the valence shell up to pqn 10, plus a few excitations from inner shells. Transition cross sections and rates are obtained from simple screened-hydrogenic formulas. Details about the atomic model construction can be found in \cite{scott_advances_2010}. The main advantage of this approach is that the atomic models are compact, computationally efficient and give reasonably accurate ionization balance over a wide range of conditions. These models have been applied to a number of high energy density physics applications. For example, non-LTE opacities produced from these models are a critical component in large scale radiation hydrodynamic simulations of laser holhraum experiments at the National Ignition Facility (NIF). \cite{rosen_role_2011,jones_progress_2017}.

In the next two sections, we use the computational model described here to examine the effects of atomic kinetics on two fundamental problems relevant to laser-produced plasmas: IB heating and non-local thermal transport.

\section{Inverse Bremsstrahlung Heating}
\label{sec:ib}
In this section, we consider IB heating and ionization of a uniform plasma due to a laser. In this case, the transport term in Eq. (\ref{eq:vbfp}) is neglected, and the electromagnetic acceleration term due to the laser is replaced by the so-called Langdon operator\cite{langdon_nonlinear_1980}. With these assumptions, we only need to solve for the isotropic part of the electron distribution function $ f_0$:
\begin{align}
\label{eq:bfp}
\partial_t f_0 (v,t) = \mathcal{C}_0 ( f_0) + \mathcal{Q}_0 ( f_0) + \mathcal{I} ( f_0)
\end{align}
where $\mathcal{I}$ denotes the Langdon operator (see appendix \ref{app:ib}). This operator was first derived by Langdon\cite{langdon_nonlinear_1980} to study non-linear IB heating and absorption. Langdon showed that during the heating, the distribution can deviate from a Maxwellian and take a super-Gaussian form of order $m>2$. As a consequence, the absorption rate is reduced compared the Maxwellian value. He proposed a fit for the ratio of the absorption rate over the Maxwellian value as follows:
\begin{equation}
\label{eq:abs}
R(\alpha) = 1-0.553/ \left[ 1+ \left( 0.27/\alpha \right)^{0.75} \right]
\end{equation}
where $R$ depends only on $\alpha$. The Langdon parameter $\alpha$ is defined by $\alpha = \frac{Z v_0^2}{v_t^2}$, where $v_0$ is the oscillation velocity of an electron in the laser electric field and $v_t \equiv \sqrt{\frac{T_e}{m_e}}$ is the thermal velocity. Here, the plasma ionization state is defined as $Z = \sum_i z_i^2 y_i / \sum_i z_i y_i$ where $z_i$ is the charge number of atomic state $i$. It can be seen that $\alpha$ represents the ratio of the thermalization to the IB heating timescale. In the limit of $\alpha \rightarrow 0$, $R\rightarrow 1$ and the distribution becomes Maxwellian. Matte et al.\cite{matte_non-maxwellian_1988} performed kinetic simulations of both uniform and non-uniform plasmas to study the effect of IB heating on the shape of the distribution function, and provided a practical formula to fit the distribution. The distribution is fit to a super Gaussian, i.e., $f \propto \exp \left[ -\left( v/v_m\right)^m \right]$ where $v_m^2 / v_t^2 = 3\varGamma (3/m) / \varGamma (5/m)  $. The parameter $m$ can be determined from:
\begin{align}
\label{eq:matte}
m (\alpha) = 2+3/\left(1 + 1.66/\alpha^{0.724} \right)
\end{align}
Weng et al.\cite{weng_inverse_2009} proposed a modified form of the Langdon operator extending its applicability to high laser intensities, where the original operator gives inaccurate results. We note that none of these authors included the effects of atomic kinetics in their calculations. In the following simulations, we demonstrate that atomic kinetics can modify the shape of the distribution function during the IB heating process, and hence affecting the absorption rate. For simplicity, we use the original Langdon operator instead of the one proposed by Weng et al. Our numerical simulations confirm that in the absence of atomic kinetics, Eqs. (\ref{eq:abs}) and (\ref{eq:matte}) provide very good fits to the calculated absorption rates and electron distributions.

\begin{figure}
\includegraphics{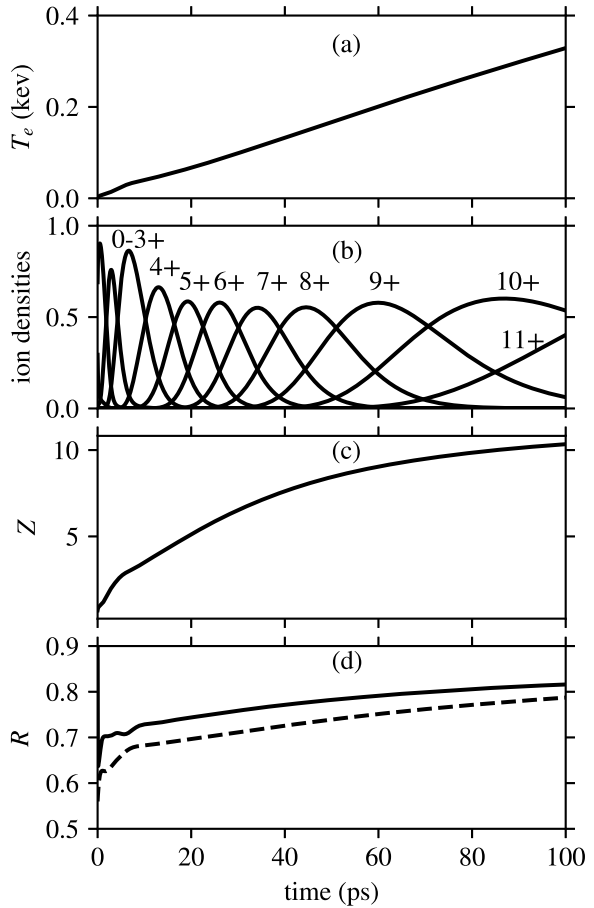}
\caption{Simulation of IB heating and ionization of Al at $n_i = 10^{19}$ cm$^{-3}$. The plasma is initially in LTE at 2 ev. (a) Electron temperature, (b) charge state densities, (c) ionization state and (d) IB absorption coefficient are shown as functions of time. In (d), solid line refers to the absorption coefficient from the simulation, while the dashed line is calculated from Eq. (\ref{eq:abs}) using $\alpha$ from the simulation.}
\label{fig:al}
\end{figure}

\begin{figure}
\includegraphics[scale=1]{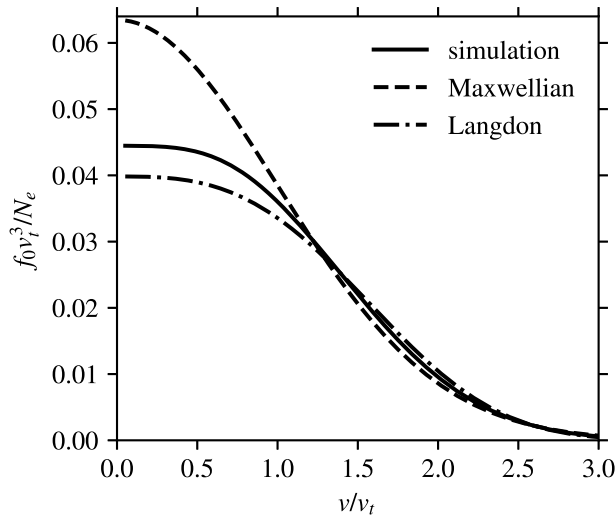}
\caption{Electron distribution function at 1.3 ps. The solid line is the distribution from simulation, and the dashed and dash-dotted lines are Maxwellian and Langdon distributions respectively. The Langdon distribution is a super Gaussian distribution of order $m=2.96$, where $m$ is determined from Eq. (\ref{eq:matte}) using $\alpha$ from the simulation. The two analytical distributions are normalized to have the same density and temperature as the actual distribution.}
\label{fig:al_dist}
\end{figure}

We simulate an Aluminum plasma of density $10^{19}$ cm$^{-3}$ being heated by a 3$\omega$ laser ($\lambda_L = 351$ nm) at a constant intensity of $10^{14}$ W/cm$^2$ for 100 ps. The plasma is initially in LTE at 2 ev corresponding to $Z \simeq 0.7$. The time evolution of several quantities of interest is shown in Fig. \ref{fig:al}. The plasma is rapidly heated to approximately 330 ev and ionized to $Z \simeq 10$ within 100 ps as shown in Fig. \ref{fig:al}a-c. Here, the electron temperature is defined as $T_e = \frac{4\pi}{3} \int_0^\infty f_0 (v) m_e v^4 \, d v$. The ratio of the absorption coefficient to the Maxwellian value is shown in Fig. \ref{fig:al}d (solid line) and also compared to the value given by Eq. (\ref{eq:abs}) (dashed line). Similar to Langdon, we define $R=f_0(v=0)/f_M(v=0)$ where $f_M$ is the equivalent Maxwellian distribution with same density and temperature. It can be seen that the value of $R$ from the simulation is always higher than that given by Eq. (\ref{eq:abs}). This suggests that the distributions appear to be more thermalized when atomic kinetics is included. To examine this further, the electron distribution at 1.3 ps (where the difference in $R$ is the largest) is plotted in Fig. \ref{fig:al_dist}. We also show comparisons with a Maxwellian distribution (dashed line) and a Langdon distribution (dash-dotted line) at the same density and temperature. The Langdon distribution takes a super Gaussian form of order $m$, where $m$ is determined from Eq. (\ref{eq:matte}). The actual distribution with self-consistent treatment of atomic kinetics looks more thermalized, i.e., closer to a Maxwellian distribution, than a Langdon distribution. To understand this, we first note that IB heating shifts both slow ($v/v_t \lesssim 1.2$) and fast ($v/v_t \gtrsim 2.6$) particles to the intermediate range, so the Langdon distribution is flattened near $v=0$. In contrast, atomic kinetic processes, mainly collisional ionization, affect the distribution in the opposite way. Since the mean ionization energy is typically a few times the electron temperature ($\sim$ 2-3 $T_e$), electrons in the range of $1.2 \lesssim v/v_t \lesssim 2.6$ have very large collisional ionization rates (fast electrons can also ionize but their cross sections are smaller while slow electrons are not sufficiently energetic). Therefore, these electrons can participate in collisional ionization and end up at lower energies. In addition, new electrons produced as a result of ionization also tend to have distributions peaked at low energies. The net result is that the distribution is shifted to lower energies; hence the distribution looks more like a Maxwellian distribution.

\begin{figure}
\includegraphics{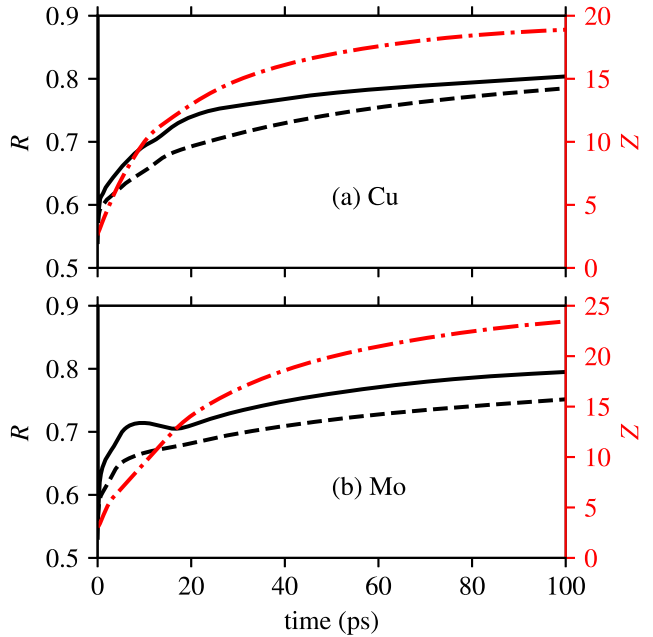}
\caption{Simulations of IB heating and ionization of (a) Cu and (b) Mo at $n_i = 10^{19}$ cm$^{-3}$. The plasma is initially in LTE at 5 ev. In both plots, the black solid and dashed lines are IB absorption coefficients $R$ and the red dash-dotted lines are the ionization state $Z$. Solid lines refer to the absorption coefficients from the simulation, while dashed lines are calculated from Eq. (\ref{eq:abs}) using $\alpha$ from the simulation.}
\label{fig:cu_and_mo}
\end{figure}

To show that this effect is not unique for Aluminum, we repeat the same simulation but using different materials. Fig. \ref{fig:cu_and_mo} shows results for similar IB heating tests using (a) Copper and (b) Molybdenum. Although the ionization process (red dash-dotted curves in Fig.~\ref{fig:cu_and_mo}) is different for each of these materials, the absorption rates (solid and dashed black curves in Fig.~\ref{fig:cu_and_mo}) are consistently higher than those predicted from Eq. (\ref{eq:abs}). In addition to modifying the absorption rates, non-Maxwellian electron distributions can also impact how various plasma parameters are inferred from measurements.~\cite{hansen_hot-electron_2002,milder_impact_2019} The results in this section indicate that atomic kinetics need to be included in the model to capture the correct distribution.

\section{Nonlocal Thermal Transport}
\label{sec:transport}
In this section, we consider a non-local thermal transport problem in a uniform plasma. Similar problems had also been explored by various authors in the past \cite{bell_elecron_1981,matte_electron_1982,albritton_nonlocal_1986}. In these works, the authors assumed a fully ionized plasma and neglected the effect of atomic kinetics. Nevertheless, they demonstrated that thermal transport in the presence of a strong temperature gradient (often as a result of laser absorption) exhibits (nonlocal) kinetic features, and therefore cannot be modeled by the standard Spitzer-Harm heat conduction model \cite{spitzer_transport_1953}. This is because the main heat carriers (electrons at $\sim 3.7 v/v_t$) can travel many mean-free-path's ahead of the thermal front and deposit heat (preheat), which leads to a reduction in the heat flux (``flux inhibition''). We will demonstrate that atomic kinetics does not alter the overall picture of nonlocal heat flow, but rather influences the propagation of the heat via the conductivity of the plasma.

The problem simulated here is identical to that presented in \cite{matte_electron_1982}, except that we also include atomic kinetics so ionization can be self-consistently modeled. A 400 $\mu$m slab of iron (Fe) plasma, initially at a density of $2 \times 10^{19}$ cm$^{-3}$ and temperature of 50 ev, is suddenly heated from the left side by a temperature source of 1 kev. This boundary condition sets up a heat wave traveling to the right. As the heat wave passes, it heats and ionizes the plasma, raising the effective ionization and heat conductivity. The ions are assumed to be static and cold. The standard treatment in hydro codes requires solving a diffusion equation for electron temperature (or energy), e.g., $\partial_t T_e = -\nabla \kappa \nabla T_e$, where the heat conductivity $\kappa$ is determined from the Spitzer-Harm (SH) formula. In this case, the heat flow is highly nonlocal so SH model does not give correct results, both in terms of heat flux and propagation of the thermal wave. For comparison purpose, we run a similar simulation using the VFP model at a fixed ionization of 18, i.e., $N_e = 3.6 \times 10^{20}$ cm$^{-3}$. Here $Z=18$ is chosen to roughly match the propagation speed of the thermal wave. The spatial domain is discretized into 200 zones and the electron velocity grid is discretized into 60 groups. We assume a constant value of $\ln \varLambda$ for both ei and ee collisions ($\ln \varLambda_{ee} = \ln \varLambda_{ei} = 7.1$).

\begin{figure}
\includegraphics{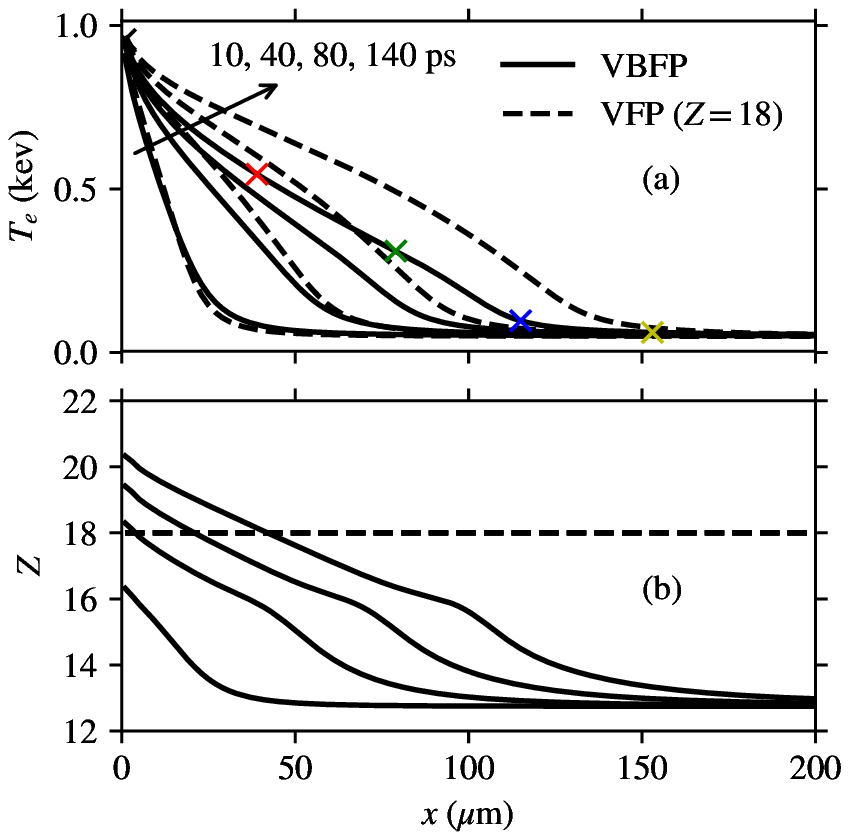}
\caption{(a) Temperature and (b) ionization state from both VFBP (solid) and VFP (dashed) simulations at four times: 10, 40, 80 and 140 ps (thermal wave moves from left to right). VBFP simulation includes self-consistent atomic kinetics, while VFP simulation assumes $Z=18$. The crosses mark spatial locations used in Fig.~\ref{fig:edf_kinetic}}
\label{fig:te_z_kinetic}
\end{figure}

\begin{figure}
\includegraphics{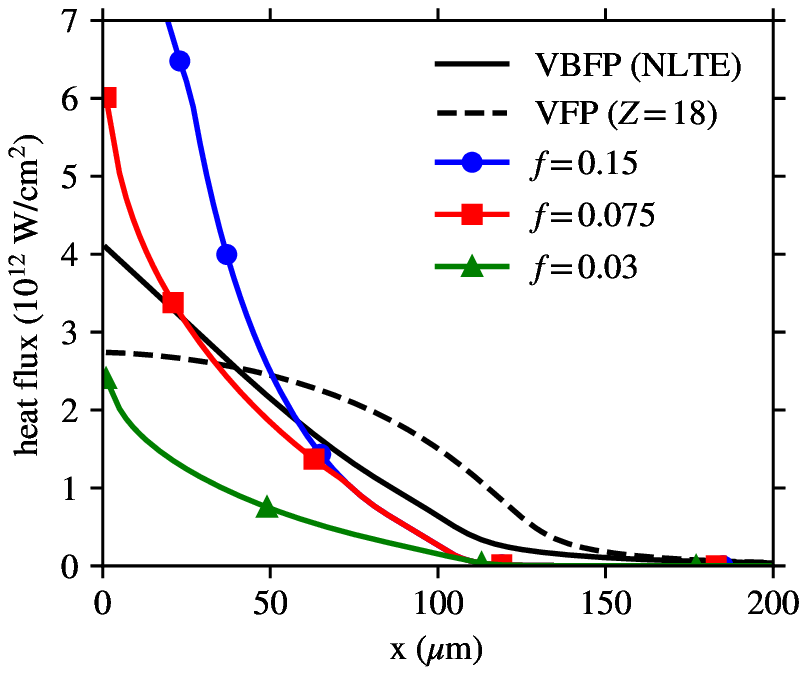}
\caption{Heat fluxes at 140 ps from VFBP and VFP simulations. Different values of flux-limited SH heat flux are also shown for comparison purpose.}
\label{fig:heatflux_kinetic}
\end{figure}

\begin{figure}
\includegraphics{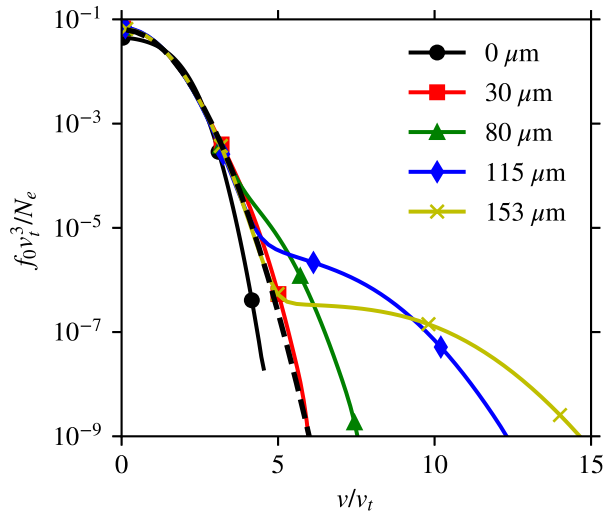}
\caption{Electron distribution functions at 140 ps and at five different locations as marked in Fig. \ref{fig:te_z_kinetic}. All distributions are normalized to have same density and energy. The dashed line is the equivalent Maxwellian distribution.}
\label{fig:edf_kinetic}
\end{figure}

Fig. \ref{fig:te_z_kinetic} show the electron temperature and ionization state at several instances in time (10, 40, 80 and 140 ps) as the heat wave travels to the right. The solid lines are solutions from the VBFP model and the dashed lines from VFP model. Up to 10 ps, the propagation and the temperature profile from VBFP and VFP are quite similar. However, at later times, the two solutions start deviating from each other. This can be understood by examining the ionization state shown Fig. \ref{fig:te_z_kinetic}b. Since the VFP simulation assumes a constant $Z$, the conductivity only depends on the temperature of the plasma (here conductivity should only be understood qualitatively as the speed at the which the heat is transported). In contrast, the VBFP simulation takes into account ionization, so the conductivity depends both on temperature and mean ionization, which vary both in space and time. To further illustrate this, Fig. \ref{fig:heatflux_kinetic} shows the heat fluxes from both simulations at 140 ps. It can be seen that the VBFP heat flux (solid line) is larger than the VFP one (dashed line) at $x \lesssim 42$ $\mu$m and smaller at $x\gtrsim 42$ $\mu$m. This exactly coincides with the ionization from VBFP simulation being higher at $x\lesssim 42$ $\mu$m and lower at $x \gtrsim 42$ $\mu$m (Fig. \ref{fig:te_z_kinetic}b). For this reason, the heat front in the VFP simulation travels faster than that in the VBFP simulation. The SH heat fluxes with different values of flux limiters are also shown in Fig. \ref{fig:heatflux_kinetic}, and clearly fail to capture the true heat flux both in magnitude and the extent of the heat flow. Fig. \ref{fig:edf_kinetic} shows the electron distribution function from the VBFP simulation at five different locations as marked in Fig. \ref{fig:te_z_kinetic}. This illustrates the origin of nonlocal heat flow as dicussed previously, that is, heat-carrying electrons travel ahead of the heat front and preheat the plasma upstream. As a result, the electron distributions upstream of the heat front show a depletion of fast particles, and the ones downstream show enhancements of fast particles. Although VFP simulations can also capture this effect, the effective conductivity (and hence the heat propagation) cannot be accurately determined without atomic kinetics. This is particularly relevant to heat flows in the corona of a high-$Z$ laser-produced plasma, e.g., a gold holhraum wall in indirect drive ICF \cite{jones_progress_2017,dewald_x-ray_2008}, where the timescales associated with atomic processes can be comparable to transport timescales.

\section{Summary}
\label{sec:summary}
We have introduced a computational model to self-consistently simulate atomic kinetics and electron kinetic physics. The model consists of a kinetic (Vlasov-Boltzmann-Fokker-Planck) equation for electrons and a collisional-radiative system of rate equations for atomic states. The electron distribution function, expanded in spherical harmonics, is self-consistently evolved in time along with the distribution of atomic state populations.

We simulated IB heating and ionization of mid- to high-Z materials to examine the influence of atomic kinetics on the heating. We found that the resultant electron distributions during the heating process always appear to be more thermalized than the classical self-similar solution of Langdon\cite{langdon_nonlinear_1980}. As a result, the absorption rates are higher than that predicted by Langdon. Non-Maxwellian distributions are also relevant for a number of experimental measurements that are sensitive to the shape of the distribution.\cite{hansen_hot-electron_2002,milder_impact_2019}

We also simulated a nonlocal thermal transport problem in a uniform plasma taking into account detailed atomic kinetics. The overall picture of nonlocal heat flow remains unchanged, but the propagation of thermal wave depends strongly on atomic kinetics. This is because the effective thermal conductivity is determined from the ionization balance, which needs to be accounted for in the model.

\appendix

\section{Inelastic collision operator}
\label{app:inelas}
In this appendix, we give mathematical expressions for all relevant processes included in the inelastic collision operator. As mentioned in the text, these operators are only applied to the isotropic component $f_0$. It is more convenient to work in energy space $\varepsilon = mv^2/2$; hence we can define an electron energy distribution function $n(\varepsilon)$ such that $n(\varepsilon) \, d\varepsilon = 4\pi f_0 (v) v^2 \, dv$. These collision terms take the form of a Boltzmann collision operator. For each type of process, we give the expression for a single transition between a single pair of atomic states. The full expression includes a sum over all transitions present in the atomic model. For each transition, we always write both forward and reverse processes, for which the cross sections are related through the principle of detailed balance.

\subsection{Excitation/Deexcitation}
The source term for an excitation/deexcitation between a pair of atomic states $i$ and $j$ is as follows:
\begin{align}
\left[ \partial_t n (\varepsilon) \right]^{ed} = \int \left( \delta_1 - \delta_0 \right) \left[ y_i n(\varepsilon_0) \sigma^{exc}_{i \rightarrow j} v_0 \right. \nonumber \\
\left. - y_j n(\varepsilon_1) \sigma^{dex}_{i \leftarrow j} v_1 \right] \, d\varepsilon_0
\label{eq:ed}
\end{align}
where we have defined $v_k = \sqrt{2 \varepsilon_k /m_e}$ and $\delta_k = \delta \left( \varepsilon - \varepsilon_k \right)$ ($k=0,1,2$). Here $\sigma^{exc}_{i \rightarrow j}$ and $\sigma^{dex}_{i \leftarrow j}$ denote excitation and deexcitation cross sections. From energy conservation, we also have $\varepsilon_0 = \varepsilon_1 + \varepsilon_{ij}$ where $\varepsilon_{ij}$ is the transition energy. The $\delta$-function's allow us to write Eq. (\ref{eq:ed}) in a compact form.

\subsection{Ionization/Recombination}
The source term for an ionization/recombination between two atomic states $i$ and $j$, where $j$ belongs to the ionized stage, reads:
\begin{align}
\label{eq:ir}
\left[ \partial_t n (\varepsilon) \right]^{ir}= \int \left( -\delta_0 +\delta_1 + \delta_2 \right) \left[ y_i n(\varepsilon_0) v_0 \sigma^{ion}_{i \rightarrow j} \right. \nonumber \\
\left. - y_j n(\varepsilon_1) n(\varepsilon_2) v_1 v_2 \sigma^{rec}_{i \leftarrow j}\right] \, d\varepsilon_1 \, d\varepsilon_2
\end{align}
where $\sigma^{ion}_{i \rightarrow j}$ and $\sigma^{rec}_{i \leftarrow j}$ denote ionization and recombination cross sections, respectively. From energy conservation, we have $\varepsilon_0 = \varepsilon_1 + \varepsilon_2 + \varepsilon_{ij}$. Since every collisional ionization produces an additional electron, Eq. (\ref{eq:ir}) involves a double integral over the energies of the scattered and ionized electrons. The three-body recombination integral is quadratic in $n (\varepsilon)$ since this process requires two free electrons. For this reason, this term constitutes the most computationally intensive part of the calculation. Most of the data for collisional ionization cross sections is given in terms of total cross sections, i.e., integrated over all transferred energy. To determine $\sigma^{ion}_{i \rightarrow j}$, we assume a Rutherford cross section, i.e., $\sigma^{ion }_{i \rightarrow j} (\varepsilon) \propto 1/(\varepsilon W^2)$ ($W$ is the energy transfered durding the collision), and scale it to get to the correct total cross section. A detailed treatment of this can be found in \cite{le_conservative_2017}.

\subsection{Autoionization/Electron Capture}
For an autoionization/electron capture between $i$ and $j$, the source term is as follows:
\begin{align}
\left[ \partial_t n (\varepsilon) \right]^{aug} = \delta (\varepsilon - \varepsilon_{ij}) \left[ y_i R^{au}_{i \rightarrow j} - y_j n(\varepsilon) v \sigma^{ec}_{i \leftarrow j} \right]
\end{align}
where $R^{au}_{i \rightarrow j}$ is the autoionization rate, and $\sigma^{ec}_{i \leftarrow j}$ the electron capture cross section. This term does not involve an integral over energy because this process only produces/removes electrons at the transition energy $\varepsilon_{ij}$.

\subsection{Photoionization/Radiative recombination}
The source term for photoionization/radiative recombination between $i$ and $j$ is as follows:
\begin{align}
\label{eq:pi}
\left[ \partial_t n (\varepsilon) \right]^{pir} = \int \left[ y_i  \frac{I_\nu}{h\nu} \sigma^{pi}_{i \rightarrow j}  - y_j n(\varepsilon) v \sigma^{rr}_{i \leftarrow j} \right. \nonumber \\
\left. \left( 1 + \frac{c^2}{2h\nu^3} 	I_\nu \right) \right] \, d\varOmega
\end{align}
where $I_\nu$ denotes the radiation intensity, and $ \sigma^{pi}_{i \rightarrow j}$ and $\sigma^{rr}_{i \leftarrow j}$ are the photoionization and radiative recombination cross sections, respectively. From energy conservation, we have $h\nu = \varepsilon + \varepsilon_{ij}$. The first term in the square bracket is due to photoionization and the second term is due to radiative recombination, which includes both spontaneous and stimulated terms. Since we assume that the plasma is optically thin, we simplify Eq. (\ref{eq:pi}) by neglecting terms depending on the radiation field:
\begin{align}
\left[ \partial_t n (\varepsilon) \right]^{rr} = - 4\pi y_j n(\varepsilon) v \sigma^{rr}_{i \leftarrow j}
\end{align}

\section{Langdon Operator}
The Langdon operator describing IB heating~\cite{langdon_nonlinear_1980} is written as:
\label{app:ib}
\begin{align}
\label{eq:ib}
\mathcal{I} (f_0) \equiv \left[ \partial_t f_0 (v) \right]^{IB} = \frac{Av_0^2}{v^2} \frac{\partial }{\partial v} \left( \frac{g(v)}{v} \frac{\partial f_0}{\partial v} \right)
\end{align}
where $A = 2\pi Z e^4 N_e \ln \varLambda_{ei}$ and $g(v) \simeq 1$. The IB heating and absorption rate can be obtained by taking a second moment of Eq. (\ref{eq:ib}), i.e., $\partial_t E_t =2\pi \int_0^\infty \mathcal{I}(f_0) \, mv^4 \, dv$. It can be easily shown that $ \partial_t E_t \propto f_0(v=0)$, so the ratio of the absorption rate to the Maxwellian value can be determined as $R = f_0(v=0) / f_M(v=0)$.


\section*{Acknowledgement}
This work was performed under the auspices of the U.S. Department of Energy by Lawrence Livermore National Laboratory under Contract DE-AC52-07NA27344.

This document was prepared as an account of work sponsored by an agency of the United States government. Neither the United States government nor Lawrence Livermore National Security, LLC, nor any of their employees makes any warranty, expressed or implied, or assumes any legal liability or responsibility for the accuracy, completeness, or usefulness of any information, apparatus, product, or process disclosed, or represents that its use would not infringe privately owned rights. Reference herein to any specific commercial product, process, or service by trade name, trademark, manufacturer, or otherwise does not necessarily constitute or imply its endorsement, recommendation, or favoring by the United States government or Lawrence Livermore National Security, LLC. The views and opinions of authors expressed herein do not necessarily state or reflect those of the United States government or Lawrence Livermore National Security, LLC, and shall not be used for advertising or product endorsement purposes.

\newpage

\end{document}